# Interplay between Charge-Density-Wave, Superconductivity, and Ferromagnetism in $CuIr_{2-x}Cr_xTe_4$ Chalcogenides

*Lingyong Zeng[a], Xunwu Hu[b], Ningning Wang[c], Jianping Sun[c], Pengtao Yang[c], Mebrouka Boubeche[a], Shaojuan Luo[d], Yiyi He[a], Jinguang Cheng[c], Dao-Xin Yao[b], Huixia Luo[a]**

[a]School of Materials Science and Engineering, State Key Laboratory of Optoelectronic Materials and Technologies, Key Lab of Polymer Composite & Functional Materials, Guangzhou Key Laboratory of Flexible Electronic Materials and Wearable Devices, Sun Yat-Sen University, No. 135, Xingang Xi Road, Guangzhou, 510275, P. R. China

[b]School of Physics, Center for Neutron Science and Technology, State Key Laboratory of Optoelectronic Materials and Technologies, Sun Yat-Sen University, Guangzhou, 510275, China

[c]Beijing National Laboratory for Condensed Matter Physics and Institute of Physics, Chinese Academy of Sciences and School of Physical Sciences, University of Chinese Academy of Sciences, Beijing 100190, China.

[d]School of Chemical Engineering and Light Industry, Guangdong University of Technology, Guangzhou, 510006, P. R. China.

ABSTRACT We report the crystal structure, charge-density-wave (CDW), superconductivity (SC), and ferromagnetism (FM) in $CuIr_{2-x}Cr_xTe_4$ ($0 \leq x \leq 2$) chalcogenides. Powder x-ray diffraction (PXRD) results reveal that $CuIr_{2-x}Cr_xTe_4$ series are distinguished between two structural types and three different regions: (i) layered trigonal structure region, (ii) mixed phase regions, (iii) spinel structure region. Besides, Cr substitution for Ir site results in rich physical properties including the collapse of CDW, the formation of dome-shaped like SC, and the emergence of magnetism. Cr doping slightly elevates the superconducting critical temperature ($T_{sc}$) to its highest $T_{sc}$ = 2.9 K around $x$ = 0.06. As $x$ increases from 0.3 to 0.4, the ferromagnetic Curie temperature ($T_c$) increases from 175 to 260 K. However, the $T_c$ remains unchanged in the spinel range of $1.9 \leq x \leq 2$. This finding provides a comprehensive material platform for investigating the interplay between CDW, SC, and FM multipartite quantum states.

**KEYWORDS** Superconductivity; Charge-density-wave; Ferromagnetism; Phase transition; First-principles calculations.

The subtle balance of different interactions among spins, phonons, and electrons gives rise to intriguing macroscopic quantum phenomena such as colossal magnetoresistance (CMR), superconductivity (SC), and multiferroicity.[1,2] Studying the interaction between these various degrees of freedom in strongly coupled lattice-electron systems is thus essential for understanding them and further improving their properties.[3] SC has long been one of the frontier topics in condensed matter physics. In particular, the discovery of magnetic superconductors has been raised a fascinating issue concerning the origin of SC.[4-6] A fascinating class of strongly correlated materials is the unconventional superconductors, which includes the materials with the range of superconducting critical temperature $T_{sc}$, from organic and heavy-fermion superconductors with relatively low $T_{sc}$ to iron pnictides and cuprates that can record $T_{sc}$ > 100 K.[7-11] Generally, in strongly correlated electron systems, SC appears around phase boundaries of magnetic phases by chemical doping or adding physical pressure. However, only a few compounds are found coexistence of SC and FM or AFM ordering, such as UMGe (M = Rh, Co),[12,13] $Sm_{1-x}R_xNiC_2$ (R = La, Lu),[14] $EuFe_2(As_{1-x}P_x)_2$,[15] $Fe_{1-x}Pd_xTe$,[16] $Ce_3PdIn_{11}$,[17] $Er_2O_2Bi$,[18] $(Li_{0.8}Fe_{0.2})OHFeSe$[19] etc. A crucial matter that could be clarified it is experimentally finding the relationship between SC and magnetism. Besides, there is no denying that prominent achievements have been realized in the experimental study, whereas the superconducting mechanism of the unconventional superconductors with a broad array of highly correlated electron systems is still a challenging job for theoretical scientists.

On the other hand, there is a common phenomenon in the conventional low dimensional layered transition-metal dichalcogenides (TMDs), at which a dome-shape like superconducting phase is usually linked with the suppression of charge density wave (CDW).[20-26] This behavior is very similar to that in the aforementioned unconventional superconductors. Thus, it gives rise to the enthusiastic study on the interaction between CDW and SC and the possible clue between this kind of TMD superconductors and high-temperature (high-$T_{sc}$) superconductors.

Recently, the coexistence of CDW and SC transition is observed in the quasi two-dimensional (2D) layered $CuIr_2Te_4$ compound, where the electronic states observed around the Fermi level mostly derive from Ir *d* and Te *p* orbitals.[27] In addition, the CDW order can be destroyed by hole doping (e.g., $CuIr_{2-x}Ti_xTe_4$, $CuIr_{2-x}Ru_xTe_4$) or electron substitution (e.g., $Cu_{1-x}Zn_xIr_2Te_4$), where dome-shape like diagrams have been observed.[28-30] Nevertheless, the CDW transition first vanishes and possibly remerges in the $CuIr_2Te_{4-x}I_x$ and $CuIr_2Te_{4-x}Se_x$ cases due to the possible disorder effect.[31,32] On the other hand, the study of the impurity effect (such as Mn, Zn, Ni, Cr) on superconducting properties prevails in the high-$T_{sc}$ superconductors, where a determination of the impurity effect is crucial for establishing the microscopic origin of SC in these materials.[33-38] Especially, most of the cases in the high-$T_{sc}$ superconductors, the Cr doping suppresses SC via impurity effect.[38] Besides, $CuCr_2Te_4$ has been previously demonstrated to be a typical spinel (space group *Fd-3m*, No. 227) ferromagnetic metal with a Curie temperature $T_c \approx 326$ K.[39-43] Thus, it will be interesting in exploring the magnetic Cr impurity effect on the physical and structural properties of layered $CuIr_2Te_4$ compound.

To this end, $CuIr_{2-x}Cr_xTe_4$ alloys were prepared successfully via a solid-state method. The crystal structures of $CuIr_{2-x}Cr_xTe_4$ alloys can be distinguished by two structural types and three regions: (i) trigonal structure with *P-3m1* space group, (ii) mixed phase; (iii) spinel type with *Fd-3m* space group. Strikingly, a complex electronic phase diagram is formed, where the CDW quickly vanishes with tiny Cr doping and $T_{sc}$ is enhanced to 2.9 K at $x = 0.06$, while SC disappears and ferromagnetic order is induced at $x = 0.3$ and remains at the whole spinel region from 1.9 to 2. Therefore, the phase diagram of $CuIr_{2-x}Cr_xTe_4$ shows multiple electronic orders where the SC phase exists between CDW and FM. To the best of our current knowledge, the SC competing with CDW and FM in one system is very rare yet valuable despite the fact that quantum critical behavior is not inapparent.

Polycrystalline samples of $CuIr_{2-x}Cr_xTe_4$ were synthesized via physical solid-state method using Cu powder (99%), Ir powder (99.9%), Cr powder (99.5%), and Te powder (99.999%) as raw materials. Stoichiometric mixtures of raw powders with 5 at% excess of Te was mixed and sealed in an evacuated silica tube, and heated in a box furnace. The heat treatment condition depends on Cr doping concentration. For $CuIr_{2-x}Cr_xTe_4$ ($0 \le x \le 0.4$), the silica tube sintered at 850 $^{o}$C for 120 h at the rate of 1 ºC/min. For $CuIr_{2-x}Cr_xTe_4$ ($1.9 \leq x \leq 2$), the heat treatment condition was 500 ºC for 120 h with a heating rate 1 ºC/min. After first heating, the obtained samples were reground and pelletized again, then heated up in the same procedure to improve the sample homogeneity. PXRD data were obtained by MiniFlex, Rigaku with Cu *Kα1* radiation. Rietveld refinements were performed with the FullProf suite software. The electrical resistivity and heat capacity were carried out using a Physical Property Measurement System (Quantum Design). The magnetic susceptibility was measured by a Quantum Design Magnetic Property Measurement System (MPMS).

We performed First-principles calculations using projector augmented wave method[44] using the Vienna ab initio simulation package (VASP).[45,46] Generalized gradient approximation (GGA) described by Perdew-Burke-Ernzerhof[47] functional was used for electron exchange correlation. The kinetic cutoff energy was set to 400 eV and Γ-centered Monkhorst-Pack scheme of 3×7×3 k-points were sampled for the Brillouin zone integration.[48] In order to consider the doping effect, lattice constants, and atomic positions were fully relaxed until the Hellmann-Feynman forces on each atom were less than 0.01 eV/Å. Zero damping DFT-D3 correction method of Grimme was considered in all the calculations.[49]

All PXRD data are refined through Rietveld method using FullProf suite software for clarifying the purity and structure of the samples. **Figure. S1** shows the detailed refinement results and crystal structures of selected $CuIr_{1.8}Cr_{0.2}Te_4$, $CuIr_{1.7}Cr_{0.3}Te_4$, and $CuCr_2Te_4$ compounds. In these cases, a good agreement between the calculated and observed PXRD patterns is found, which verifies the validity of the employed structural models. And the refinement results of the representative samples $CuIr_{1.8}Cr_{0.2}Te_4$, $CuIr_{1.7}Cr_{0.3}Te_4$, and $CuCr_2Te_4$ are presented in **Table S1.** **Figure. 1a,c** displays the PXRD patterns of all the polycrystalline $CuIr_{2-x}Cr_xTe_4$ samples collected at room temperature. PXRD patterns show different features between $0 \leq x \leq 0.4$ and $1.9 \leq x \leq 2$ doping ranges. It indicates that the pristine $CuIr_2Te_4$ forms a trigonal structure, belonging to the *P-3m1* space group. The samples remain in the same trigonal space group in the doping range of $0 \leq x \leq 0.4$, though tiny Ir impurity is present, and the *a* and *c* lattice parameters both increase with increasing Cr doping concentration. For the doping region of $0.4 < x < 1.9$, we find that the spinel $CuCr_2Te_4$ phase appears except for the layered tetragonal phase, indicating they are two-phase mixtures. When $x \geq 1.9$, the main phase changes into a spinel type with space group *Fd-3m*, and the lattice parameters also increase with increasing *x*. Nevertheless, a small amount of $CuTe_2$ is observed as an impurity, which is hard to be eliminated through various attempts. Similar impurity peaks are also observed in the previous literature.[42] For the end member of $CuCr_2Te_4$, the refined lattice parameters are $a = b = c = 11.137$ Å, which are consistent with the previous report.[42] With the Cr doping increasing, the XRD peaks shift systematically (**Figure. 1a**), suggesting that most Cr are incorporated into the lattices. The high-resolution TEM images of $CuIr_{1.94}Cr_{0.06}Te_4$ polycrystalline powder directly visualize the lattice fringes and confirm the crystalline nature (see **Figure. S2a**). We further measured SEM-EDXS characterization of $CuIr_{2-x}Cr_xTe_4$ ($x$ = 0.30, 0.35, 0.40) for checking the homogeneity and true ratio of the samples. As shown in **Figure. S3** and **S4** from supporting information, all constituent elements distribute homogeneously. However, we can see that the obtained samples are be more Cu-rich and Te-deficient compared with those of the target compounds, which may be caused by a small amount of Te evaporation during synthesis process (see **Table. S2**).

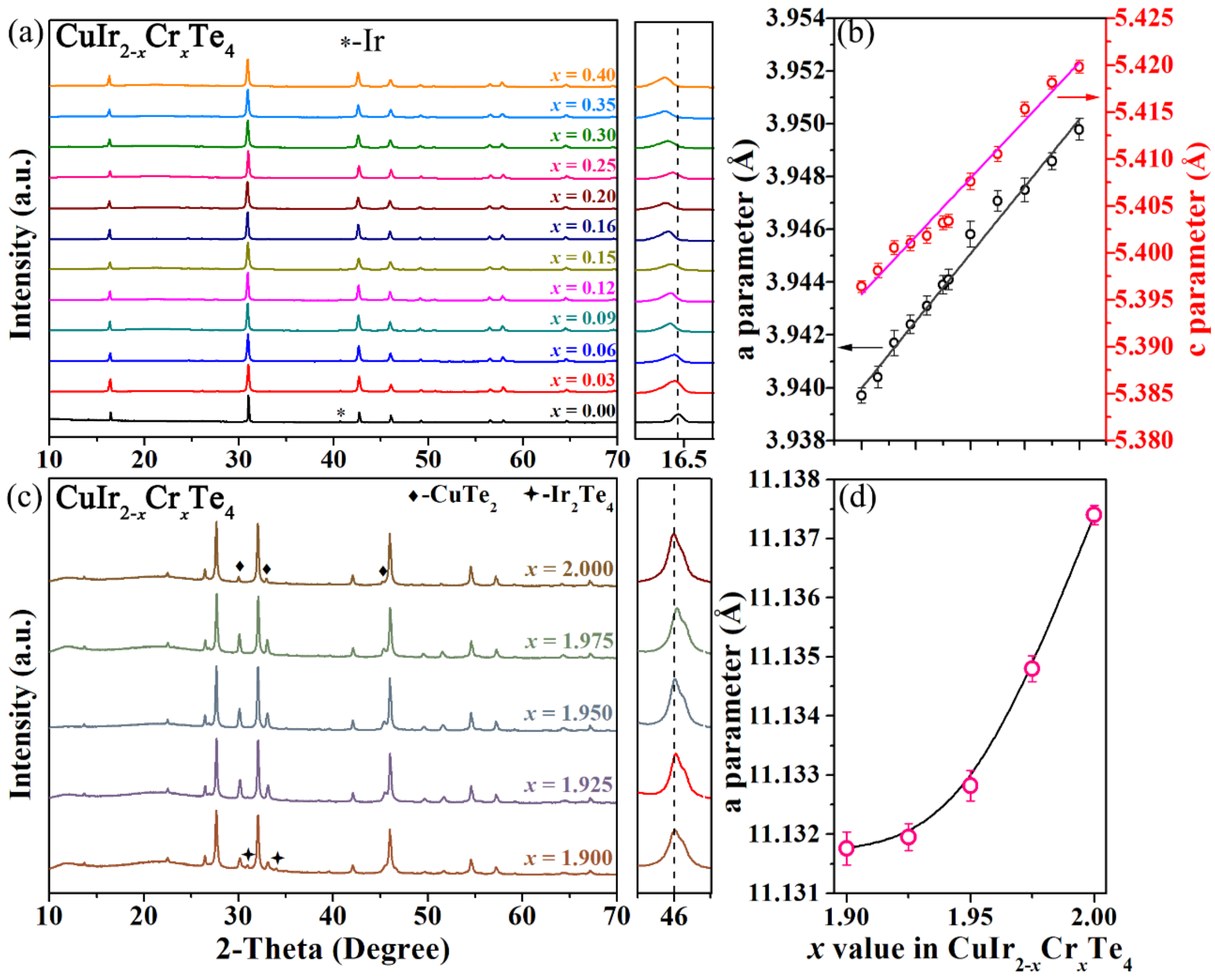

**Figure. 1** (a) Powder XRD patterns of $CuIr_{2-x}Cr_xTe_4$ ($0 \le x \le 0.4$) samples. (b) Cr content dependence of the lattice parameters of $CuIr_{2-x}Cr_xTe_4$ ($0 \le x \le 0.4$). (c) Powder XRD patterns of $CuIr_{2-x}Cr_xTe_4$ ($1.9 \le x \le 2$) samples. (d) Cr content dependence of lattice parameters as a function of Cr content $x$ of $CuIr_{2-x}Cr_xTe_4$ ($1.9 \le x \le 2$).

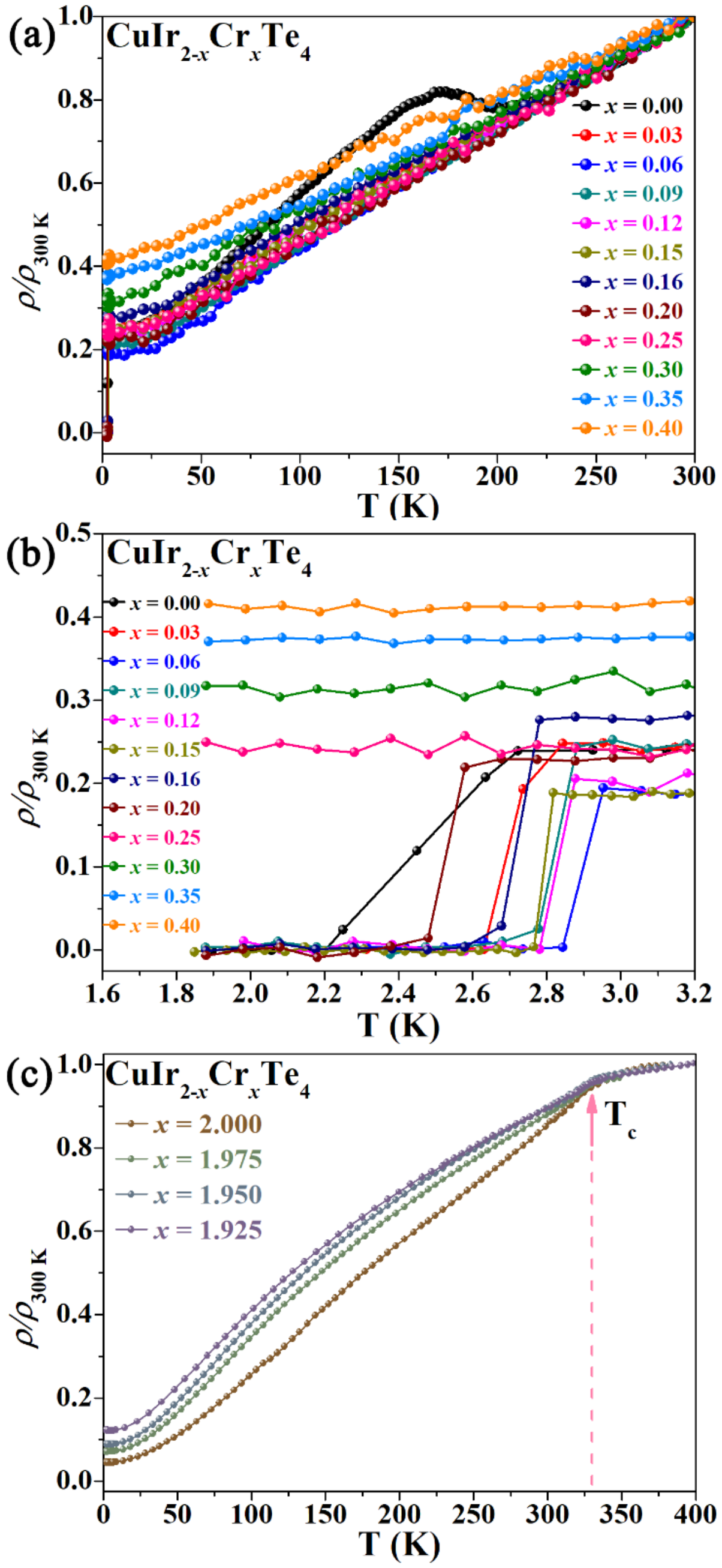

**Figure. 2** (a) Normalized temperature-dependent resistivity of $CuIr_{2-x}Cr_xTe_4$ ($0 \leq x \leq 0.4$) over the temperature range of 300 - 1.8 K. (b) Normalized temperature-dependent resistivity of $CuIr_{2-x}Cr_xTe_4$ ($0 \leq x \leq 0.4$) at low temperature. (c) Normalized temperature-dependent resistivity of $CuIr_{2-x}Cr_xTe_4$ ($1.9 < x \leq 2$).

Resistivity measurements cooling from 300 K to 1.8 K were performed in the polycrystalline $CuIr_{2-x}Cr_xTe_4$ samples. For $CuIr_{2-x}Cr_xTe_4$ ($0 < x \leq 0.2$), the resistivity data exhibit metallic behavior in the normal state and undergo superconducting transitions below 3 K (**Figure. 2a**). An anomalous hump link in the CDW order around 180 K is observed in the parent $CuIr_2Te_4$. Yet, upon Cr doping, the CDW order is suppressed and vanishes in all doping samples. The rapid suppression of CDW is often observed in the presence of magnetic impurities that act as negative chemical pressure. The detailed superconducting transitions in low-temperature resistivities are depicted in **Figure. 2b**. $T_{sc}$ first increases steadily up to 2.9 K around $x$ = 0.06, and decreases gradually for $x \geq 0.09$ with increasing Cr content, and SC eventually disappears when $x > 0.25$. Compared with the parent $CuIr_2Te_4$, the doping of Cr makes the superconducting transition steeper. The normal-state resistivity of high doping spinel samples ($1.9 < x \leq 2$) shows a change of slope around $T_c$ (see **Figure. 2c**), which is related to the magnetic transition.

**Table 1.** Comparison of the superconducting parameters of $CuIr_2Te_4$ based systems.

| **Compound** | $\mathbf{CuIr_{1.94}Cr_{0.06}Te_4}$ | $\mathbf{CuIr_{1.95}Ru_{0.05}Te_4}$ **Ref [28]** | $\mathbf{CuIr_2Te_{3.9}I_{0.1}}$ **Ref [31]** | $\mathbf{CuIr_2Te_4}$ **Ref [27]** |
|---|---|---|---|---|
| $T_{sc}$ **(K)** | 2.90 | 2.79 | 2.95 | 2.50 |
| $\gamma$ **(mJ mol$^{-1}$ K$^{-2}$)** | 10.28 | 12.26 | 12.97 | 12.05 |
| $\beta$ **(mJ mol$^{-1}$ K$^{-4}$)** | 2.1 | 1.87 | 3.03 | 1.97 |
| $\Theta_D$ **(K)** | 186 | 194 | 165 | 190 |
| $\Delta C_{el.}/\gamma T_{sc}$ | 1.66 | 1.51 | 1.46 | 1.50 |
| $\lambda_{ep}$ | 0.62 | 0.65 | 0.70 | 0.63 |
| $N(E_F)$ **(states/eV f.u)** | 2.70 | 3.15 | 3.24 | 3.10 |
| $\mu_0H_{c2}$**(T) (WHH)** | 0.218 | 0.247 | 0.188 | 0.12 |
| $\mu_0H^P$**(T)** | | 5.24 | 5.543 | 4.65 |
| $\mu_0H_{c1}(T)$ | 0.018 | 0.098 | 0.024 | 0.028 |
| $\xi_{GL}$**(0) (nm)** | 43.4 | 36.3 | 41.9 | 52.8 |

To further verify the superconducting property of our samples, specific heat measurements are conducted. **Figure. 3a** shows the $C_p$(T)/T curves of $CuIr_{1.94}Cr_{0.06}Te_4$ down to 1.8 K at zero field and 1 T field. In 0 T, we observe a clear anomaly with the midpoint of superconducting transition around 2.8 K, which signifies that SC is bulk in nature. The specific heat jump was suppressed completely in applied magnetic field 1 T, indicating $\mu_0H$ = 1 T higher than the upper critical field. The normal state specific heat can be characterized by $C_p(T) = C_{el.}(T) + C_{ph.}(T)$ with electronic contribution of $C_{el.}(T) = \gamma T$ and the phonon contribution of $C_{ph.}(T) = \beta T^3$. The linear fit to the $C_p/T$ between $T_{sc}$ and 10 K, yields $\gamma = 10.28$ mJ mol$^{-1}$ K$^{-2}$ and $\beta = 2.1$ mJ mol$^{-1}$ K$^{-4}$. Further, the ratio of specific heat jump and normal state specific heat, $\Delta C_{el.}/\gamma T_{SC} = 1.66$, is higher than 1.43 expected of BCS weak-coupling limit, demonstrating it is a moderate strong coupling superconductor.[50] The Debye temperature can be obtained following $\Theta_D = (12\pi^4 nR/5\beta)^{1/3}$, where $n$ is the number of atoms in the formula unit, and $R$ is the gas constant. The calculated value of $\Theta_D$ is 186 K, which is slightly smaller than that of the undoped sample $CuIr_2Te_4$ (190 K). In the light of McMillan's theory,[51] the electron-phonon coupling constant $\lambda_{ep}$ is described with $\lambda_{ep}=\frac{1.04+\mu^* \ln\left(\frac{\Theta_D}{1.45T_{sc}}\right)}{(1-0.62\mu^*)\ln\left(\frac{\Theta_D}{1.45T_{sc}}\right)-1.04}$, where $\mu^*$ is the Coulomb pseudopotential parameter. It can estimate $\lambda_{ep} \approx 0.62$, which suggests that $CuIr_{1.94}Cr_{0.06}Te_4$ is a moderately coupling superconductor [52]. With the value of $\gamma$ and $\lambda_{ep}$, the density of states (DOSs) at Fermi level $N(E_F)$ can be obtained with the formula $N(E_F) = \frac{3}{\pi^2 k_B^2 (1+\lambda_{ep})}\gamma$ . **Table 1** summarizes the related superconducting performances of the $CuIr_2Te_4$ doping series and undoped $CuIr_2Te_4$. Compared with other optimal doping compositions with nonmagnetic dopants (e.g., Ru, I), $CuIr_{1.94}Cr_{0.06}Te_4$ shows enhanced or comparable $T_c$. Low temperature specific heat data for $CuIr_{1.94}Cr_{0.06}Te_4$ in several magnetic fields up to 0.04 T is plotted as $C_p/T$ vs. $T^2$ in **Figure. 3b**. With increasing applied field, the specific heat jump is progressively decreased.

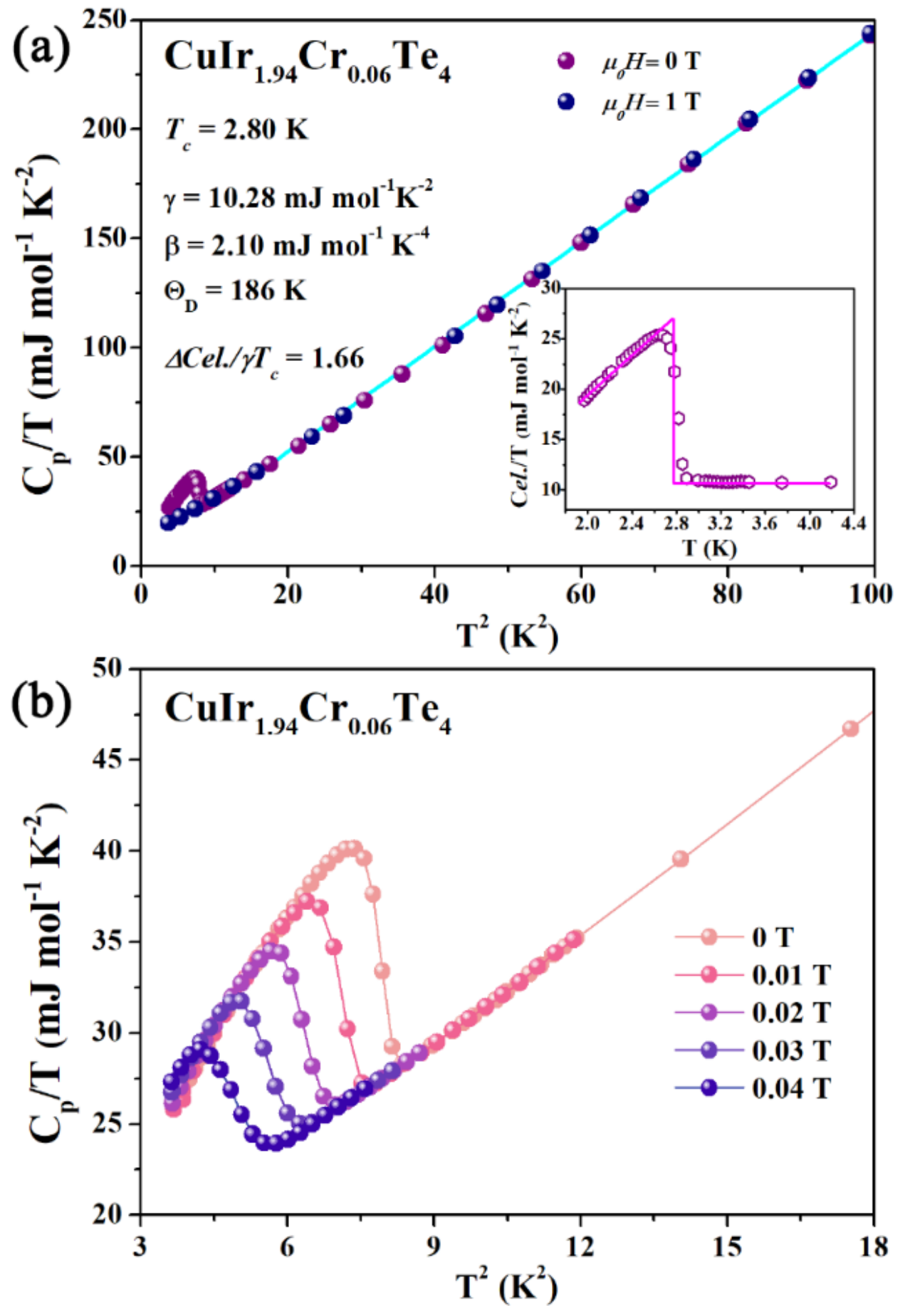

**Figure. 3** (color online) (a) The $C_p$/T vs. $T^2$ curve for $CuIr_{1.94}Cr_{0.06}Te_4$ under 0 T and 1 T. (b) Temperature dependence of specific heat divided by temperature $C_{el.}$/T of $CuIr_{1.94}Cr_{0.06}Te_4$ sample under 0 T.

To gain further insights into the superconducting state, the temperature dependent electrical resistivity of $CuIr_{1.94}Cr_{0.06}Te_4$ was tested by applying a magnetic field up to 0.08 T (see **Figure. S5a**). The zero-resistance point $T_{sc}$ gradually shifts to 1.8 K with an increasing external magnetic field. In order to calculate the upper critical magnetic field ($\mu_0H_{c2}(0)$), the critical temperature is defined by the drop of the resistivity to 10 %, 50 % and 90 % of its normal state value varies with the magnetic field in a manner displayed in **Figure. S5b**. Dot-dash lines show the fits to the data using the dirty limit Werthamer-Helfand-Hohenburg (WHH) theory, $\mu_0H_{c2}(0)$= -0.693$T_{sc}$ $\left(\frac{dH_{c2}}{dT}\right)_{T_{sc}}$, where $dH_{c2}/dT$ is the slope of $\mu_0H_{c2}$(T) near $T_{sc}$.[35] And we estimate the $\mu_0H_{c2}(0)$ to be 0.170 T for $CuIr_{1.94}Cr_{0.06}Te_4$ from 50 % $\rho_N$ criteria. The experimental $H_{c2}(T)$ data in the whole temperature range also adopt the Ginzburg-Landau (GL) formula $\mu_0 H_{c2}(T) = \mu_0 H_{c2}(0) * \frac{1-(T/T_{sc})^2}{1+(T/T_{sc})^2}$, to estimate $\mu_0 H_{c2}(0)$ = 0.218 T from 50% $\rho_N$ criteria (see solid blue line in Fig. 4(d)). Utilizing the relation $\mu_0 H_{c2} = \Phi_0/2\pi\xi^2$, where $\Phi_0$ is the flux quantum (*h/2e*), we obtain the coherence length $\xi$ = 43.4 nm. This value is smaller than that of the parent sample $CuIr_2Te_4$. Moreover, the $\mu_0 H_{c2}$s for the $CuIr_{1.91}Cr_{0.09}Te_4$ and $CuIr_{1.88}Cr_{0.12}Te_4$ samples are also calculated, as shown in **Figure. S5 d** and **f**.

**Figure. 4a** exhibits the temperature dependent DC magnetic susceptibility ($\chi(T)$) of the $CuIr_{2-x}Cr_xTe_4$ ($0 \leq x \leq 0.2$) down to 1.8 K, under conditions of zero field cooled (ZFC) at 30 Oe. The strong diamagnetic signals in the $4\pi\chi$ curves were observed below $T_{SC}^{\chi}$. The obtained $T_{SC}^{\chi}$ first increases with Cr content, and then decreases with further increasing Cr content, which is in good agreement with the SC observed in resistivity measurements (**Figure. 2b**). In addition, a large Meissner fraction close to 100 % at 1.8 K is observed in $CuIr_{1.94}Cr_{0.06}Te_4$, which rules out the possibility of filamentary SC or impurity phases. To further study the superconducting properties, we perform the magnetization measurements at low temperatures between 1.9 and 2.5 K for the $CuIr_{1.94}Cr_{0.06}Te_4$ sample. **Figure. 4b** presents the isothermal magnetization curves at different temperatures. The absolute value of magnetization increases linearly, then decreases after embracing the lower critical field ($H_{c1}$), and finally becomes a paramagnetic state above the upper critical field ($H_{c2}$), which exhibits a typical type-II superconductor behaviour. The lower critical field, $H_{c1}$, as a function of temperature, is presented in **Figure. 4d**. The lower critical fields are approximated by determining the point at which the M(H) data first deviate from linearity in the low field magnetization loops at each temperature (see **Figure. 4c**). The solid purple line represents a fit to the data using a simple parabolic T dependence in the form of $\mu_0 H_{c1}(T) = \mu_0 H_{c1}(0)(1-(T/T_{sc})^2)$. The extrapolated lower critical field ($\mu_0 H_{c1}(0)$) at 0 K is about 0.018 T.

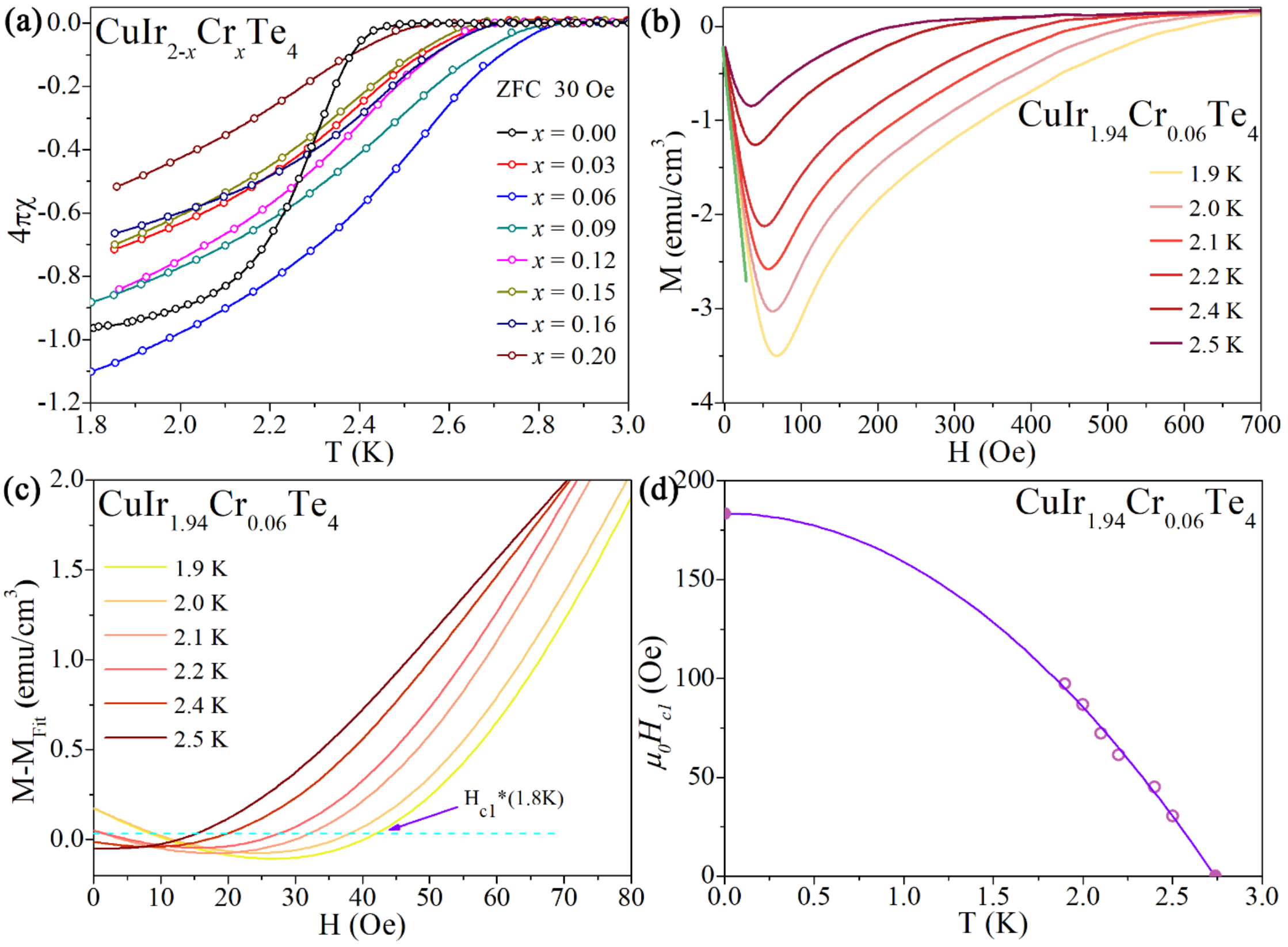

**Figure. 4** Superconducting properties examined from magnetic susceptibilities of $CuIr_{2-x}Cr_xTe_4$ ($0 \le x \le 0.2$) samples. (a) Temperature dependence of magnetic susceptibilities ($4\pi\chi$) of $CuIr_{2-x}Cr_xTe_4$ ($0 \le x \le 0.2$) samples under applied magnetic field 30 Oe. (b) M(H) curves of $CuIr_{1.94}Cr_{0.06}Te_4$ under different temperatures. (c) M-$M_{Fit}$ as a function of H. (d) The low critical field of $CuIr_{1.94}Cr_{0.06}Te_4$.

**Figure 5a** displays the temperature dependent magnetization of $CuIr_{2-x}Cr_xTe_4$ ($x$ = 0.30, 0.35, 0.40) under a fixed magnetic field of 1 T. In $CuIr_{1.7}Cr_{0.3}Te_4$, for example, while lowering the temperature, the magnetization value of $CuIr_{1.7}Cr_{0.3}Te_4$ is very low until 200 K. As exhibited in **Figure. 5b**, a sharp paramagnetic (PM) to FM phase transition arises at $T_c \sim 155$ K defined by the minimum in temperature dependence of the differential quotient of magnetization (dM/dT ~ T). Below 200 K, the magnetization increases rapidly, which may suggest the presence of ferromagnetic transition. The Curie temperature is determined to be 175 K from the Curie-Weiss plots (slanting line in **Figure. 5c**), which is almost similar to the phase transition temperature (**Figure. 5a**). With increasing $x$, the phase transition temperature of PM to FM moves to a higher temperature. When the temperature is further reduced, the magnetization rises sharply below 20 K, suggesting the appearance of Pauli paramagnetic region. At high temperature above 250 K, $\chi$(T) can be well fitted using the modified Curie-Weiss law $\chi = \chi_0 + C/(T-\theta)$, where $\chi_0$ is the temperature independent contribution to magnetic susceptibility, $\theta$ is the Weiss temperature and $C$ is the Curie constant. In $CuIr_{1.7}Cr_{0.3}Te_4$, for example, we obtain the values of $\chi_0 = 4.16 \times 10^{-4}$ emu mol$^{-1}$ per f.u.$^{-1}$ by fitting the inverse susceptibility data (250 - 350 K), the Curie constant $C$ is 0.233 and the Curie-Weiss constants $\theta$ = 169.8 K. Positive Curie-Weiss constant indicate the dominance of ferromagnetic interactions in $CuIr_{1.7}Cr_{0.3}Te_4$. **Figure. 5d** shows the magnetic susceptibility for $CuIr_{2-x}Cr_xTe_4$ ($x$ = 0.30, 0.35, 0.4) as a function of temperature from 400 - 700 K under 1 T magnetic field. In $CuIr_{1.7}Cr_{0.3}Te_4$, for example, the data is almost linear and the slope can be obtained to be 1.526 by linear fitting ($\frac{1}{\chi-\chi_0} \sim \frac{1}{C_{f.u.}} T$). Using the per formula unit fit, $C_{f.u.}$= 1/slope = 0.655. According to the formula $\mu_{eff} = \sqrt{\frac{8}{n} \times C_{f.u.}}$, $n$ is the number of Cr ions per formula unit. The effective moment $\mu_{eff}$ for $CuIr_{1.7}Cr_{0.3}Te_4$ is obtained to be 4.18 $\mu_B$ Cr-ion$^{-1}$. Considering the Cu-rich and Te-deficient for some polycrystalline samples (see Table S2), the effective moment $\mu_{eff}$ for the real $Cu_{1.13}Ir_{1.73}Cr_{0.31}Te_{3.83}$ composition is 4.11 $\mu_B$ Cr-ion$^{-1}$, which is close to the spin only value 3.87 $\mu_B$ expected for S = 3/2 ($Cr^{3+}$ ion).[42] The effective moment for $CuIr_{1.65}Cr_{0.35}Te_4$ and $CuIr_{1.6}Cr_{0.4}Te_4$ are 4.02 and 4.07 $\mu_B$, respectively. Based on the EDXS results, the effective moment for the real composition $Cu_{1.09}Ir_{1.65}Cr_{0.36}Te_{3.90}$ and $Cu_{1.15}Ir_{1.61}Cr_{0.41}Te_{3.83}$ are 3.96 and 4.02 $\mu_B$, respectively. There is no big difference between the designed and obtained compositions. It is noted that the effective moment per Cr is observed to be only weakly dependent on Cr concentration (see inset of **Figure. 5c**). This phenomenon is also observed in $Ti_{1-x}Cr_xSe_4$, in which the effective moment is ~ 4 $\mu_B$, and is basically not affected by the change of Cr content.[53] In the meantime, the magnetic susceptibility of $CuIr_{2-x}Cr_xTe_4$ ($0 < x \leq 0.2$) roughly follows Curie-Weiss law under the normal state without any magnetic phase transition.

**Figure. 5e** shows the temperature dependent magnetization under an applied field 1 T of $CuIr_{2-x}Cr_xTe_4$ ($x$ = 1.9, 1.925, 1.95, 1.975, 2) samples. The inset displays the temperature derivative of M, dM/dT, as a function of the temperature. The negative peak temperature of dM/dT is consistent with the change slope of resistivity measurements in **Figure. 2c**. The inverse magnetic susceptibility in an applied magnetic field 1 T can be also fitted using the modified Curie-Weiss law $\chi = \chi_0 + C/(T-\theta)$. In $CuIr_{0.05}Cr_{1.95}Te_4$, for example, from the fit to the inverse susceptibility data, we obtain the values of $\chi_0$ = -0.01611 emu $mol^{-1}$ per $f.u.^{-1}$, the Curie constant $C$ is 5.10. The effective moment $\mu_{eff}$ is calculated to be 3.99 $\mu_B$ Cr-$ion^{-1}$, which is near the effective moment 4.14 $\mu_B$ of $CuCr_2Te_4$ previously reported.[42]

Hysteresis loops for $CuIr_{2-x}Cr_xTe_4$ ($x$ = 0.30, 0.35, 0.40) samples under various temperatures up to H = ± 7 T are shown in **Figure. S6**, which displays the typical behaviour of ferromagnetic materials. Ferromagnetism is frequently observed in Cr intermetallic compounds and Cr-doped compounds, such as for the two-dimensional (2D) ferromagnets $Cr_2Ge_2Te_6$ [54-56] and Cr-doped layered dichalcogenide $TiSe_2$.[53] The M-H curves are essentially linear at $T > T_c$, indicating a simple paramagnetic state. The saturated moment at 2 K for $CuIr_{1.7}Cr_{0.3}Te_4$, $CuIr_{1.65}Cr_{0.35}Te_4$, and $CuIr_{1.6}Cr_{0.4}Te_4$ are 0.60, 0.87, and 0.67 $\mu_B$ Cr-$ion^{-1}$, respectively. The value of saturated moment is much less than effective moment. This phenomenon has also been observed in other Cr-based materials such as $Cr_4PtGa_{17}$,[57] $CrBe_{12}$,[58] and $CuCr_2Te_4$.[42] The magnetization is not saturated at 7 T, the possible mechanism may be a non-collinear alignment.[42,59]

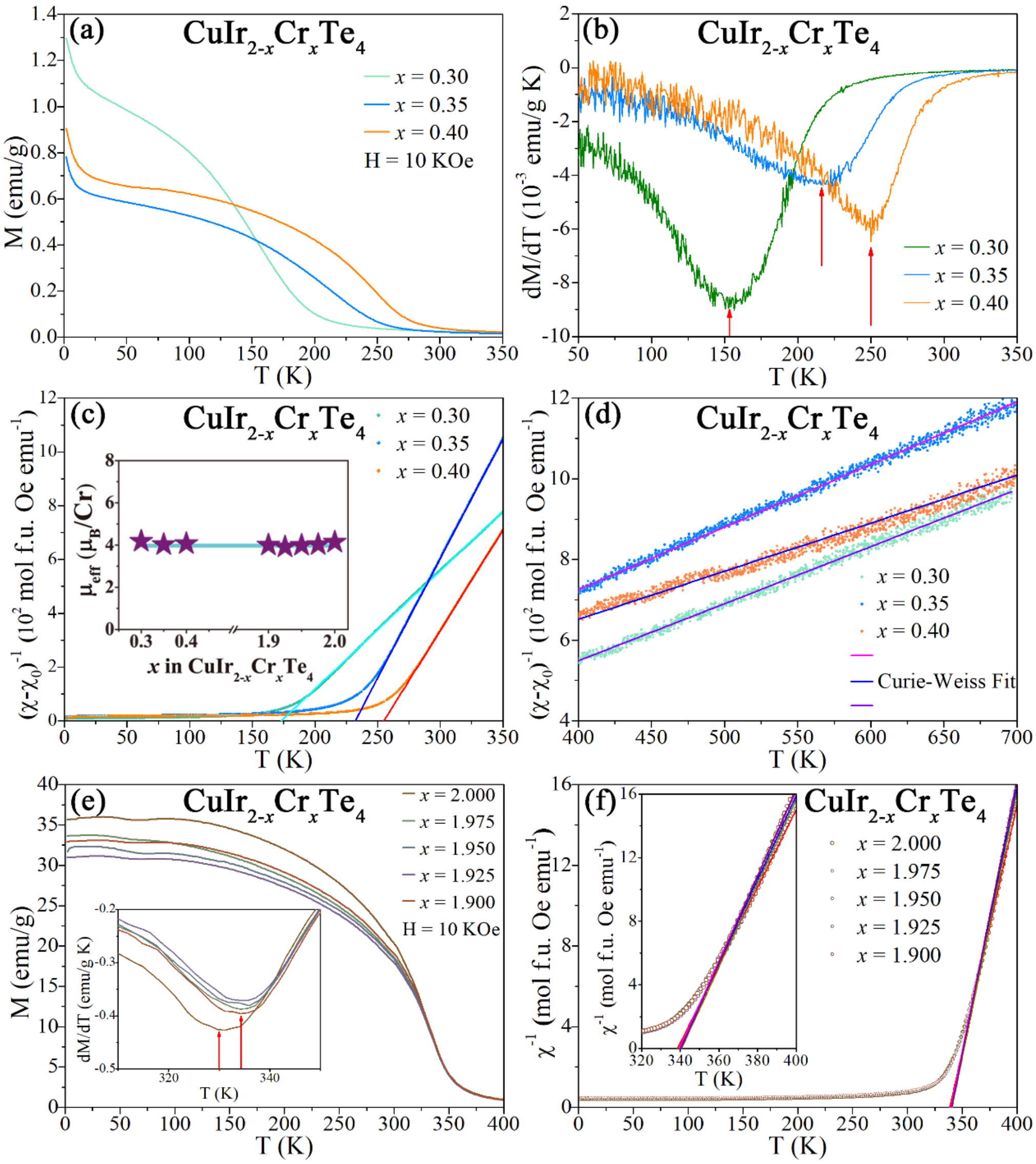

**Figure. 5** (a) Temperature dependence of magnetization M for $CuIr_{2-x}Cr_xTe_4$ ($x$ = 0.30, 0.35, 0.40) samples at H = 10 KOe. (b) The derivative (dM/dT) curves of $CuIr_{2-x}Cr_xTe_4$ ($x$ = 0.30, 0.35, 0.40) samples. (c) Magnetic susceptibility of $CuIr_{2-x}Cr_xTe_4$ ($x$ = 0.30, 0.35, 0.40), which is fitted by the modified Curie-Weiss law. The inset shows the effective moment per Cr of $CuIr_{2-x}Cr_xTe_4$ samples. (d) The temperature dependence of inverse magnetic susceptibility $(\chi - \chi_0)^{-1}(T)$ at T > 400 K for $CuIr_{1.7}Cr_{0.3}Te_4$. (e) Temperature dependence of magnetization M for $CuIr_{2-x}Cr_xTe_4$ ($x$ = 1.925, 1.95, 1.975, 2) samples at H = 10 KOe. The inset shows the derivative (dM/dT) curves. (f) Magnetic susceptibility of $CuIr_{2-x}Cr_xTe_4$ ($x$ = 1.925, 1.95, 1.975, 2), which is fitted with the Curie-Weiss law.

Based on the above results, the substitution of Ir by Cr results in an extremely rich phase diagram for $CuIr_{2-x}Cr_xTe_4$ ($0 \leq x \leq 2$), which is displayed in **Figure. 6**. XRD measurements indicate that there exists doping induced structural transition from a trigonal phase at $0 \leq x \leq 0.4$ to a spinel phase at $1.9 \leq x \leq 2$. The phase diagram shows the interplay between the CDW order ($T_{CDW}$), superconducting transition temperature ($T_{sc}$), and Curie temperature ($T_c$). At $x$ increases, the first order CDW transition temperature $T_{CDW}$ = 250 K can be suppressed quickly by Cr doping and $T_{sc}$ gradually increases, which is the most natural explanation for the fierce competition between CDW states and SC. The SC phase reveals a classic dome-shaped $T_{sc}(x)$ with a maximum value of 2.9 K taking place around $x$ = 0.06. As further increases Cr content, the SC vanishes around $x$ = 0.2 and induces a ferromagnetic phase at $x$ = 0.3, indicating the superconducting pairs are destroyed by the ferromagnetism. For $0.3 \leq x \leq 0.4$, FM Curie temperature $T_c$ increase with the increase in $x$. At spinel structure region ($1.9 \leq x \leq 2$), $T_c$ reaches the maximum value $T_c$ = 334 K. The phase diagram points out that SC competes with CDW and FM in the $CuIr_{2-x}Cr_xTe_4$ system.

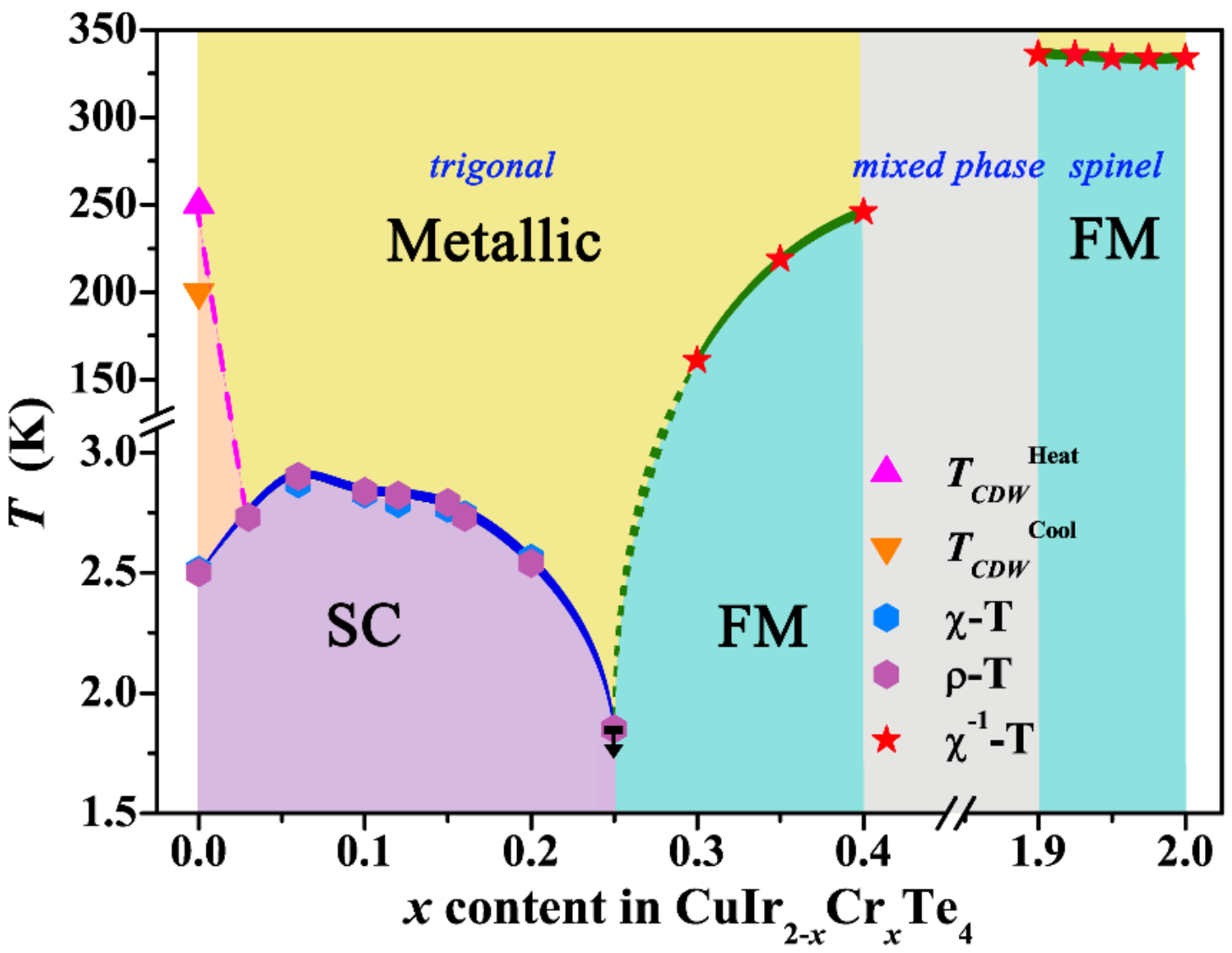

**Figure. 6** The phase diagram of $T_{CDW}$, $T_{sc}$, and $T_c$ versus Cr doping concentration in $CuIr_{2-x}Cr_xTe_4$ system.

Related to this, it is worthwhile to explore the distinct features of the phase diagram of $CuIr_{2-x}Cr_xTe_4$ system with other systems. In contrast, nonmagnetic Ru substitution at the Ir site in $CuIr_2Te_4$ also rapidly destroys the CDW state, but does not induce FM state. In addition, the CDW order is destroyed with 5*p* iodine doping at the Te site in $CuIr_2Te_4$ but unexpectedly reappears with high iodine doping concentration. Further comparison with unconventional superconductors, the competitive relationship between SC and magnetism is often observed in iron-based and copper high-temperature superconductors. Furthermore, unconventional superconductivity often arises around the point where magnetic order is suppressed by doping or using pressure to tune electronic instabilities.[60-65] Such a phase diagram with superconductivity emerging from a magnetic ground state is similar to this study. Therefore, for unconventional superconductors, an effective way to get some insight into the pairing mechanism remains to forcibly change their strength and to monitor the corresponding changes in $T_{sc}$. This is characteristically what is done in conventional superconductor $CuIr_2Te_4$ system with doping effects, where the enhancement of $T_{sc}$ with ion masses maybe related to the change of phonon frequencies. With the increase of Cr doping, CDW is quickly suppressed, at which the SC enhances and reaches its maximum value with small Cr doping. Further increasing magnetic Cr concentration, the ferromagnetic interaction force rearranges the electronic structure, thereby breaking the superconducting energy gap, thus leading to the complete suppression of SC at $x > 0.25$. Finally, Raman scattering investigations have suggested that a quantum critical point occurs within the superconducting composition area in the $Cu_xTiSe_2$ series,[66] thus a quantum critical phase transition may exist in our $CuIr_{2-x}Cr_xTe_4$ system around a critical composition $x \approx 0.25$ where the SC disappears, followed by the emergence of FM. However, it needs further study to verify it.

In order to compare with the experimental results, we further investigate the electronic structure and magnetism of the $CuIr_{2-x}Cr_xTe_4$ by combining First-principles calculations. **Figure. 7a, b, c, d** shows the density of states (DOSs) and the projected density of states (PDOSs) for each atom. The green dashed line represents the Fermi level. All the diagrams show the DOSs passing through the Fermi level, which means that the materials are gapless, indicating metallic properties. It is shown that the PDOSs of Te, Ir, and Cr elements have significant contributions in the vicinity of the Fermi level. All calculations take into account different magnetic structures. **Figure. 7a** and **Figure. 7b** show the DOSs without spin polarization at $x = 0$ and $x = 1/4$, respectively. We calculate the free energy of non-magnetic, ferromagnetic, and antiferromagnetic structures, and finally, the free energy of non-magnetic is the lowest. This means that calculations show that its ground state is non-magnetic. **Figure. 7c** and **Figure. 7d** show the DOSs with spin polarization at $x = 1/3$ and $x = 2$, respectively. The calculated results show that the ground state is ferromagnetic, the ordered magnetic moments are 2.96 $\mu_B$/Cr and 2.66 $\mu_B$/Cr, respectively. Comparative analysis based on **Figure. 6**, with the Cr doping, the $CuIr_{2-x}Cr_xTe_4$ system changes from non-magnetic to ferromagnetic. Magnetism due to Cr doping content in $CuIr_{2-x}Cr_xTe_4$ system may suppress SC and lead to structural phase transition.

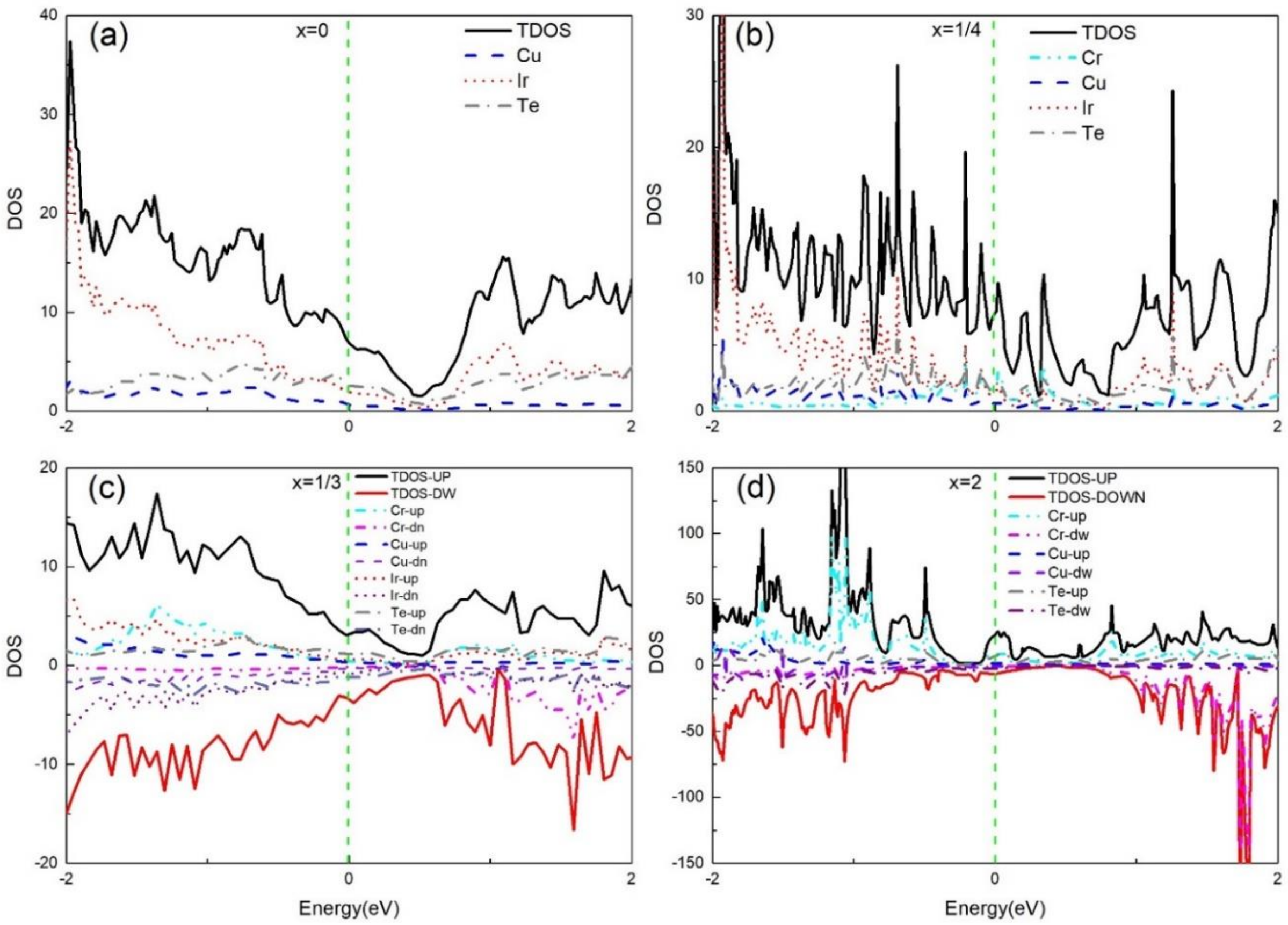

**Figure. 7** Calculated total and partial DOS for each atom in $CuIr_{2-x}Cr_xTe_4$ system. (a) $x = 0$, (b) $x = 1/4$, (c) $x = 1/3$, (d) $x = 2$.

In conclusion, the effects of Cr doping on the evolution of structure, SC, and FM in the $CuIr_{2-x}Cr_xTe_4$ system are studied. It was found that the CDW order was suppressed quickly by Cr doping, while the superconducting transition temperature $T_{sc}$ was elevated to the highest value of 2.9 K around $x = 0.06$, and SC finally disappears at $x \geq 0.25$. The ferromagnetic Curie temperature $T_c$ increases from 175 to 260 K as $x$ increases from 0.3 to 0.4. And $T_c$ remains unchanged despite the Cr doping in the range of $1.9 \leq x \leq 2$. Usually, unconventional SC appears around phase boundaries of magnetic phases in a strongly correlated electron system. As conventional superconducting materials, $CuIr_{2-x}Cr_xTe_4$ provides a rare example for studying the interplay between CDW, SC, and FM ordering. It is therefore a new platform for investigating strongly correlated phenomena, which could lead to insights into the physics of unconventional high-$T_{sc}$ superconductors and heavy fermion superconductors.

ASSOCIATED CONTENT

**Supporting Information**.

The Supporting Information is available free of charge at http://

Elemental maps and EDX (PDF)

AUTHOR INFORMATION

**Corresponding Author**

**Huixia Luo** -- School of Materials Science and Engineering, State Key Laboratory of Optoelectronic Materials and Technologies, Key Lab of Polymer Composite & Functional Materials, Guangzhou Key Laboratory of Flexible Electronic Materials and Wearable Devices, Sun Yat-Sen University, No. 135, Xingang Xi Road, Guangzhou, 510275, P. R. China.
Email: luohx7@mail.sysu.edu.cn

**Notes**

The authors declare no competing financial interest.

ACKNOWLEDGMENT

This work is supported by the NSFC-11922415, NKRDPC-2017YFA0206203, NKRDPC-2018YFA0306001, NSFC-11974432, NSFC-92165204, Guangdong Basic and Applied Basic Research Foundation (2019A1515011718, 2019A1515011337), the Fundamental Research Funds for the Central Universities (19lgzd03), the Key Research & Development Program of Guangdong Province, China (2019B110209003), Shenzhen Institute for Quantum Science and Engineering (Grant No. SIQSE202102), and the Pearl River Scholarship Program of Guangdong Province Universities and Colleges (20191001).

# Supporting Information

## Interplay between Charge-Density-Wave, Superconductivity, and Ferromagnetism in $CuIr_{2-x}Cr_xTe_4$ Chalcogenides

Lingyong Zeng[a], Xunwu Hu[b], Ningning Wang[c], Jianping Sun[c], Pengtao Yang[c], Mebrouka Boubeche[a], Shaojuan Luo[d], Yiyi He[a], Jinguang Cheng[c], Dao-Xin Yao[b], Huixia Luo[a*]

[a]School of Materials Science and Engineering, State Key Laboratory of Optoelectronic Materials and Technologies, Key Lab of Polymer Composite & Functional Materials, Guangzhou Key Laboratory of Flexible Electronic Materials and Wearable Devices, Sun Yat-Sen University, No. 135, Xingang Xi Road, Guangzhou, 510275, P. R. China

[b]School of Physics, Center for Neutron Science and Technology, State Key Laboratory of Optoelectronic Materials and Technologies, Sun Yat-Sen University, Guangzhou, 510275, China

[c]Beijing National Laboratory for Condensed Matter Physics and Institute of Physics, Chinese Academy of Sciences and School of Physical Sciences, University of Chinese Academy of Sciences, Beijing 100190, China.

[d]School of Chemical Engineering and Light Industry, Guangdong University of Technology, Guangzhou, 510006, P. R. China

Corresponding Author

* Corresponding author/authors complete details (Telephone; E-mail:) (+86)-2039386124; E-mail address: luohx7@mail.sysu.edu.cn;

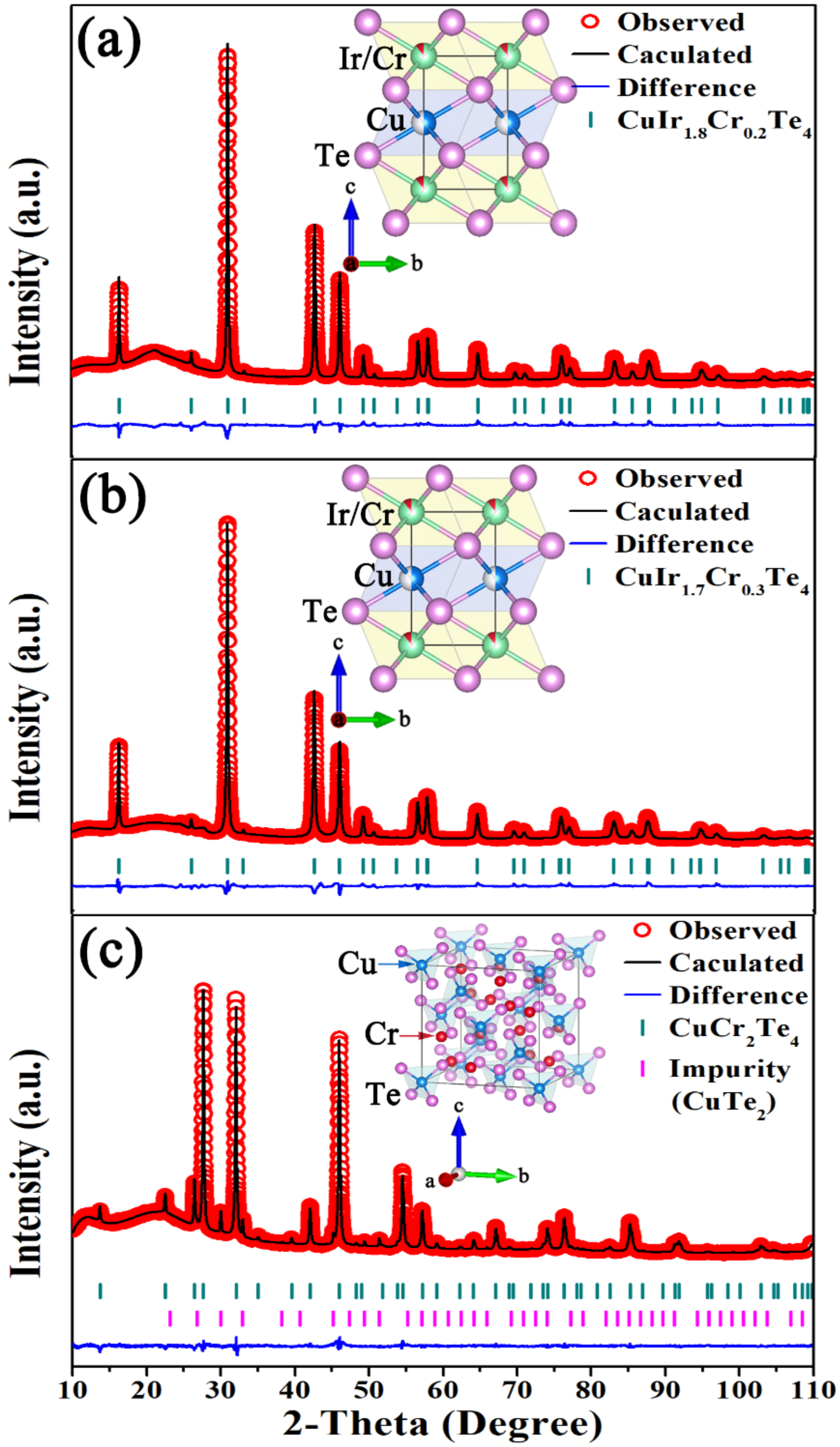

**Fig. S1** (color online) Rietveld refinement of the representative (a) $CuIr_{1.8}Cr_{0.2}Te_4$, (b) $CuIr_{1.7}Cr_{0.3}Te_4$, (c) $CuCr_2Te_4$ samples. The insets show the crystal structure of $CuIr_{2-x}Cr_xTe_4$ ($x$ = 0.2, 0.3, 2) samples.

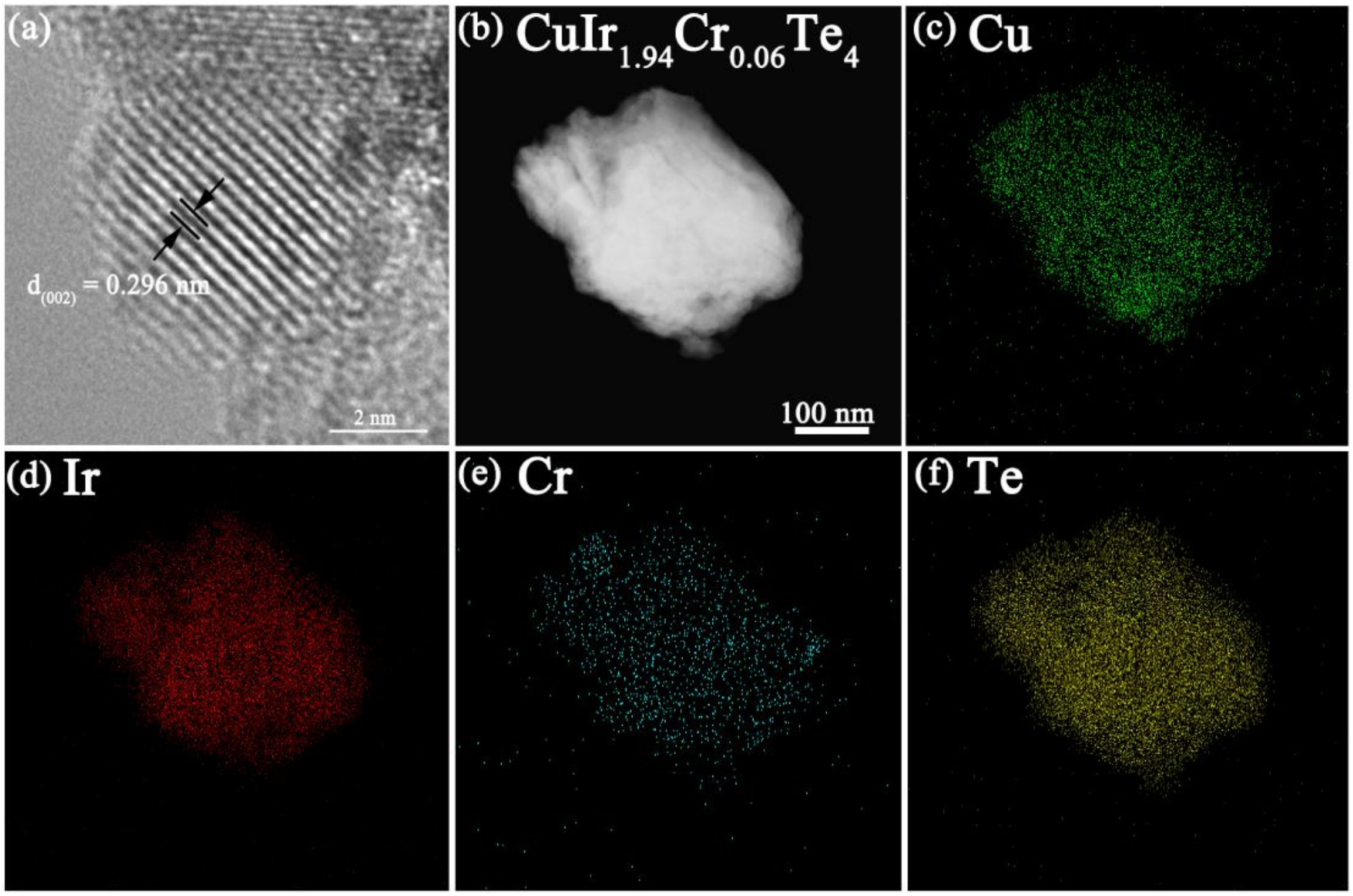

**Fig. S2** (color online) (a) HRTEM image of $CuIr_{1.94}Cr_{0.06}Te_4$ polycrystalline powder; (b) HRTEM showing where the elemental maps were obtained; (c) Cu mapping image; (d) Ir mapping image; (e) Cr mapping image; (f) Te mapping image.

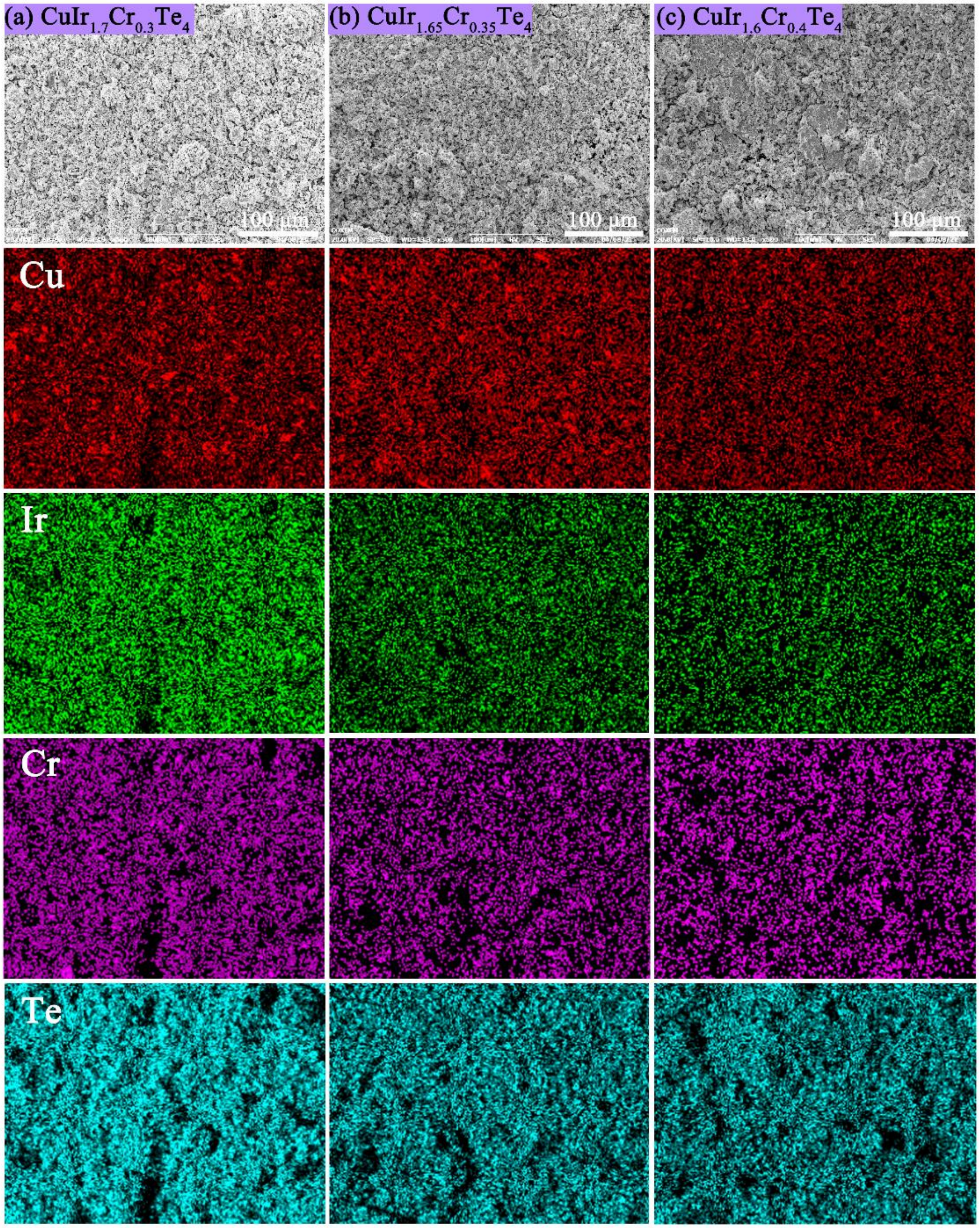

**Fig. S3** SEM and EDXS mappings of $CuIr_{2-x}Cr_xTe_4$ ($x$ = 0.30, 0.35, 0.40).

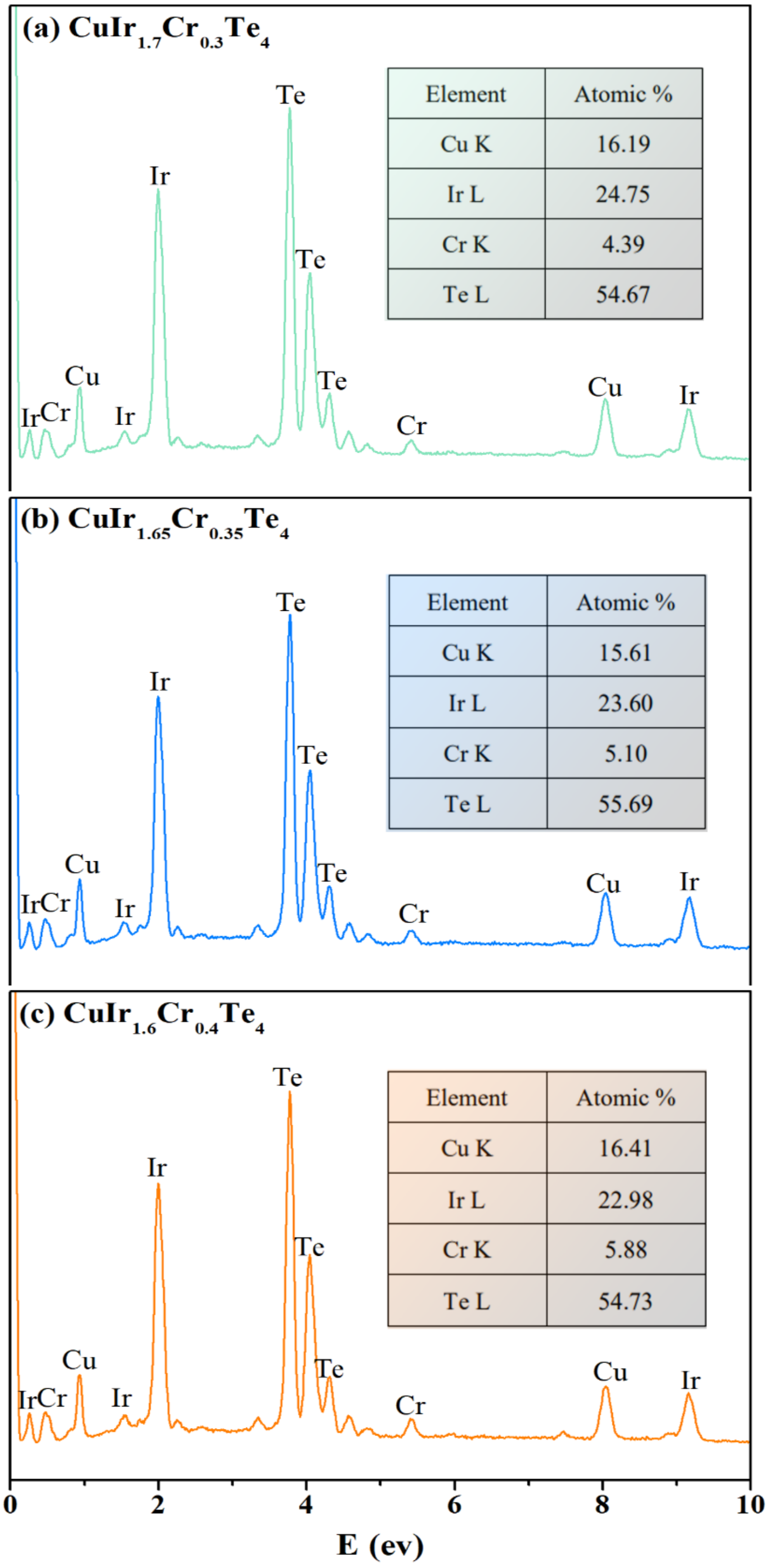

**Fig. S4** EDXS spectrum of $CuIr_{2-x}Cr_xTe_4$ ($x$ = 0.30, 0.35, 0.40).

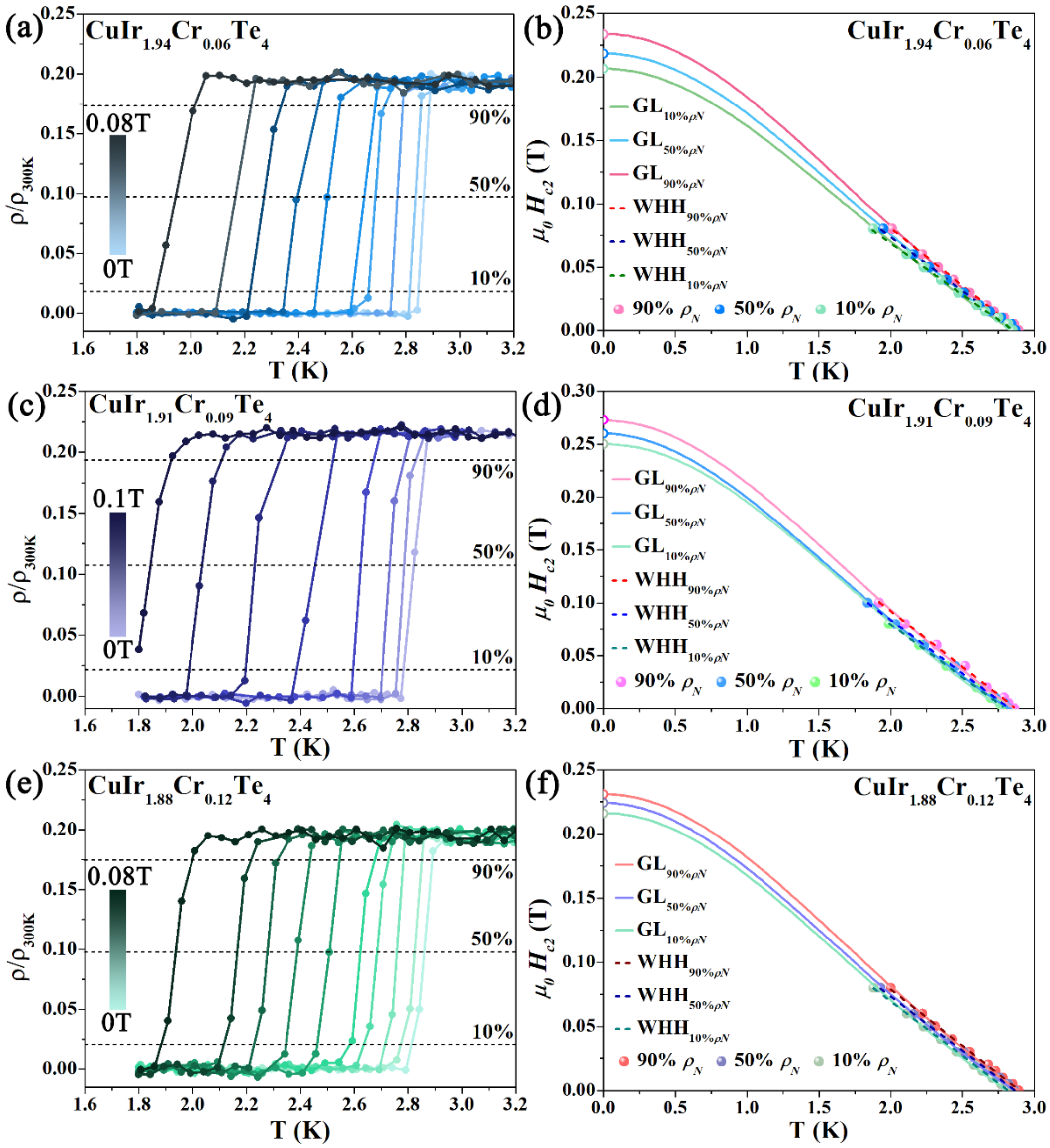

**Fig. S5** (color online) Normalized temperature-dependent resistivity of (a) $CuIr_{1.94}Cr_{0.06}Te_4$, (c) $CuIr_{1.91}Cr_{0.09}Te_4$ and (e) $CuIr_{1.88}Cr_{0.12}Te_4$ at different magnetic field. Temperature dependence of the upper critical fields of (b) $CuIr_{1.94}Cr_{0.06}Te_4$, (d) $CuIr_{1.91}Cr_{0.09}Te_4$, and (f) $CuIr_{1.88}Cr_{0.12}Te_4$.

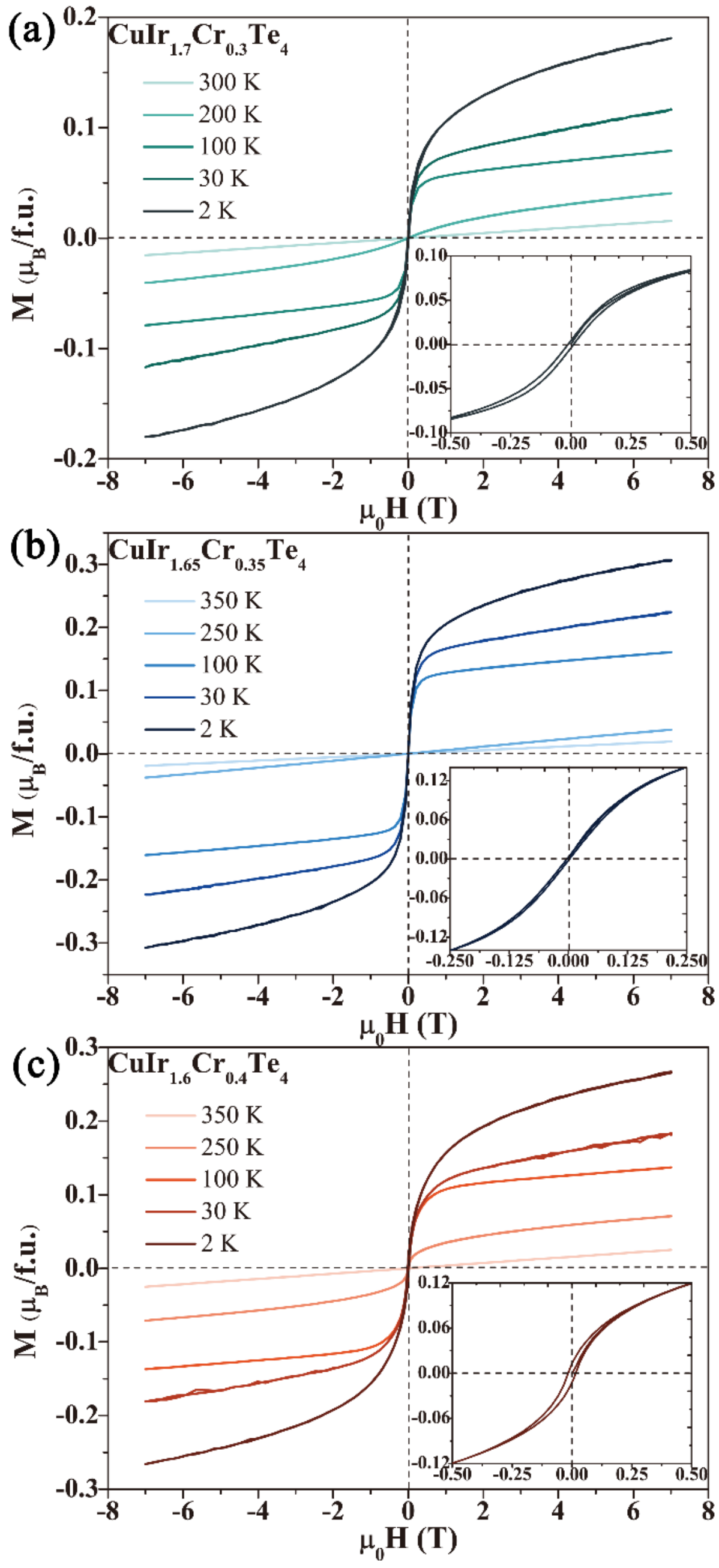

**Fig. S6** M-H curves for $CuIr_{2-x}Cr_xTe_4$ ($x$ = 0.30, 0.35, 0.40) samples under various temperatures. The insets show zoom-ins of M-H curve at 2 K.

**Table S1.** Rietveld refinement structural parameters of $CuIr_{1.8}Cr_{0.2}Te_4$, $CuIr_{1.7}Cr_{0.3}Te_4$, and $CuCr_2Te_4$ compounds

| **$CuIr_{1.8}Cr_{0.2}Te_4$** | | | $R_{wp}$ = 6.90%, $R_p$ = 4.80%, $\chi^2$ =5.84 | | | |
|---|---|---|---|---|---|---|
| **Label** | ***x*** | ***y*** | ***z*** | **site** | **Occupancy** | **Multiplicity** |
| Cu | 0 | 0 | 0.5 | 2*b* | 0.5 | 1 |
| Ir | 0 | 0 | 0 | 1*a* | 0.9 | 1 |
| Cr | 0 | 0 | 0 | 1*a* | 0.1 | 1 |
| Te | 0.33333 | 0.66667 | 0.74646 | 2*b* | 1 | 2 |
| **$CuIr_{1.7}Cr_{0.3}Te_4$** | | | $R_{wp}$ = 7.24%, $R_p$ = 5.36%, $\chi^2$ =6.88 | | | |
| **Label** | **x** | **y** | **z** | **site** | **Occupancy** | **Multiplicity** |
| Cu | 0 | 0 | 0.5 | 2*b* | 0.5 | 1 |
| Ir | 0 | 0 | 0 | 1*a* | 0.85 | 1 |
| Cr | 0 | 0 | 0 | 1*a* | 0.15 | 1 |
| Te | 0.33333 | 0.66667 | 0.74520 | 2*b* | 1 | 2 |
| **$CuCr_2Te_4$** | | | $R_{wp}$ = 2.92%, $R_p$ = 2.19%, $\chi^2$ = 1.58 | | | |
| **Label** | **x** | **y** | **z** | **site** | **Occupancy** | **Multiplicity** |
| Cu | 0 | 0 | 0 | 8*a* | 1 | 16 |
| Cr | 0.62500 | 0.62500 | 0.62500 | 16*d* | 1 | 8 |
| Te | 0.38131 | 0.38131 | 0.38131 | 32*e* | 1 | 32 |

**Table S2**. The ratio of elements for $CuIr_{2-x}Cr_xTe_4$ ($x$ = 0.3, 0.35, 0.4).

| **Element ratio / Sample** | **Cu** | **Ir** | **Cr** | **Te** |
|---|---|---|---|---|
| **$CuIr_2Te_4$ [21]** | 0.97 | 1.96 | 0 | 3.93 |
| **$CuIr_{1.7}Cr_{0.3}Te_4$** | 1.13 | 1.73 | 0.31 | 3.83 |
| **$CuIr_{1.65}Cr_{0.35}Te_4$** | 1.09 | 1.65 | 0.36 | 3.90 |
| **$CuIr_{1.6}Cr_{0.4}Te_4$** | 1.15 | 1.61 | 0.41 | 3.83 |